\documentclass[12pt]{article}
\usepackage{graphicx}
\def\nn{\noindent}
\def\Re{{\cal R \mskip-4mu \lower.1ex \hbox{\it e}\,}}
\def\Im{{\cal I \mskip-5mu \lower.1ex \hbox{\it m}\,}}
\def\ie{{\it i.e.}}
\def\eg{{\it e.g.}}

\def\etal{{\it et al.}}

\def\sub#1{_{\lower.25ex\hbox{$\scriptstyle#1$}}}
\def\tev{\,{\ifmmode\mathrm {TeV}\else TeV\fi}}
\def\gev{\,{\ifmmode\mathrm {GeV}\else GeV\fi}}
\def\mev{\,{\ifmmode\mathrm {MeV}\else MeV\fi}}
\def\mpl{\ifmmode M_{pl}\else $M_{pl}$\fi}
\def\to{\rightarrow}

\def\subw{_{\rm w}}
\def\mh{\ifmmode m\sbl H \else $m\sbl H$\fi}
\def\mch{\ifmmode m_{H^\pm} \else $m_{H^\pm}$\fi}
\def\mt{\ifmmode m_t\else $m_t$\fi}
\def\mc{\ifmmode m_c\else $m_c$\fi}
\def\mz{\ifmmode M_Z\else $M_Z$\fi}
\def\mw{\ifmmode M_W\else $M_W$\fi}
\def\mws{\ifmmode M_W^2 \else $M_W^2$\fi}
\def\mhs{\ifmmode m_H^2 \else $m_H^2$\fi}   
\def\mzs{\ifmmode M_Z^2 \else $M_Z^2$\fi}
\def\mts{\ifmmode m_t^2 \else $m_t^2$\fi}
\def\mcs{\ifmmode m_c^2 \else $m_c^2$\fi}
\def\mchs{\ifmmode m_{H^\pm}^2 \else $m_{H^\pm}^2$\fi}
\def\ztwo{\ifmmode Z_2\else $Z_2$\fi}
\def\zone{\ifmmode Z_1\else $Z_1$\fi}
\def\mtwo{\ifmmode M_2\else $M_2$\fi}
\def\mone{\ifmmode M_1\else $M_1$\fi}
\def\tb{\ifmmode \tan\beta \else $\tan\beta$\fi}
\def\xw{\ifmmode x\subw\else $x\subw$\fi}
\def\ch{\ifmmode H^\pm \else $H^\pm$\fi}
\def\lum{\ifmmode {\cal L}\else ${\cal L}$\fi}
\def\inpb{\,{\ifmmode {\mathrm {pb}}^{-1}\else ${\mathrm {pb}}^{-1}$\fi}}
\def\infb{\,{\ifmmode {\mathrm {fb}}^{-1}\else ${\mathrm {fb}}^{-1}$\fi}}
\def\epem{\ifmmode e^+e^-\else $e^+e^-$\fi}
\def\ppb{\ifmmode \bar pp\else $\bar pp$\fi}
\def\bsg{\ifmmode B\to X_s\gamma\else $B\to X_s\gamma$\fi}
\def\bsll{\ifmmode B\to X_s\ell^+\ell^-\else $B\to X_s\ell^+\ell^-$\fi}
\def\bstt{\ifmmode B\to X_s\tau^+\tau^-\else $B\to X_s\tau^+\tau^-$\fi}
\def\lamt{\ifmmode \tilde\lambda\else $\tilde\lambda$\fi}
\def\shat{\ifmmode \hat s\else $\hat s$\fi}
\def\that{\ifmmode \hat t\else $\hat t$\fi}
\def\uhat{\ifmmode \hat u\else $\hat u$\fi}

\newskip\zatskip \zatskip=0pt plus0pt minus0pt
\def\matth{\mathsurround=0pt}
\def\lsim{\mathrel{\mathpalette\atversim<}}

\def\atversim#1#2{\lower0.7ex\vbox{\baselineskip\zatskip\lineskip\zatskip
  \lineskiplimit 0pt\ialign{$\matth#1\hfil##\hfil$\crcr#2\crcr\sim\crcr}}}

\def\grtsim{\,\,\rlap{\raise 3pt\hbox{$>$}}{\lower 3pt\hbox{$\sim$}}\,\,}
\def\lsim{\,\,\rlap{\raise 3pt\hbox{$<$}}{\lower 3pt\hbox{$\sim$}}\,\,}

\renewcommand{\thefootnote}{\fnsymbol{footnote}}

\begin{document} \begin{titlepage}
\rightline{\vbox{\halign{&#\hfil\cr
&SLAC-PUB-10389\cr
&March 2004\cr}}}
\begin{center}
\thispagestyle{empty} \flushbottom \centerline{ {\Large\bf Warped
Higgsless Models with IR--Brane Kinetic Terms
\footnote{Work supported in part
by the Department of Energy, Contract DE-AC03-76SF00515}
\footnote{e-mails: $^a$hooman@ias.edu,
$^b$hewett@slac.stanford.edu, $^c$lillieb@slac.stanford.edu, and
$^d$rizzo@slac.stanford.edu}}}
\medskip
\end{center}

\centerline{H. Davoudiasl$^{1,a}$, J.L. Hewett$^{2,b}$,
B. Lillie$^{2, c}$, and T.G. Rizzo$^{2,d}$}
\vspace{8pt} \centerline{\it $^1$School of
Natural Sciences, Institute for Advanced Study,
Princeton, NJ 08540}
\vspace{8pt}
\centerline{\it $^2$Stanford Linear
Accelerator Center, Menlo Park, CA, 94025}

\vspace*{0.3cm}

\begin{abstract}
We examine a warped Higgsless $SU(2)_L\times SU(2)_R\times
U(1)_{B-L}$ model in 5--$d$ with IR(TeV)--brane kinetic terms.
It is shown that adding a brane term for the $U(1)_{B-L}$ gauge
field does not affect the scale ($\sim 2-3$ TeV) where perturbative 
unitarity in $W_L^+ W_L^- \to W_L^+ W_L^-$ is violated.  This term 
could, however, enhance the agreement of the model with the precision 
electroweak data.  In contrast, the inclusion of a kinetic
term corresponding to the $SU(2)_D$ custodial symmetry of
the theory  delays the unitarity violation in $W_L^\pm$ scattering 
to energy scales of $\sim 6-7$ TeV for a significant fraction of the
parameter space.  This is about a factor of 4 improvement compared to 
the corresponding scale of unitarity violation in the Standard Model
without a Higgs. We also show that null searches for extra gauge
bosons at the Tevatron and for contact interactions at LEP II place 
non-trivial bounds on the size of the IR-brane terms.

\end{abstract}



\renewcommand{\thefootnote}{\arabic{footnote}} \end{titlepage}


\section{Introduction}

As we enter the era of the LHC experiments, it is appropriate to
examine the features of various approaches to Electroweak
Symmetry Breaking (EWSB).  One of the latest attempts for
describing EWSB is the proposal of Refs.~\cite{flat,warped}.  In
this approach, a judiciously chosen set of boundary conditions
in a 5--$d$ Higgsless $SU(2)_L\times SU(2)_R\times
U(1)_{B-L}$ model gives rise to a pattern of gauge boson masses and 
couplings that are similar to those obtained in the Standard Model 
(SM) via a Higgs doublet condensate.
The geometry of this model is based on the Randall--Sundrum (RS) 
hierarchy solution \cite{RS}, where two branes reside at the boundaries
of a 5--$d$ Anti-de Sitter space\footnote{In
the RS background, holographic arguments based on the AdS/CFT
correspondence \cite{Maldacena} have been useful in elucidating
the features of the Higgsless theory{\cite {warped,NomuraI,Bar,NomuraII}}.}.
In this scenario, the boundary conditions give rise
to the breaking chain $SU(2)_R\times U(1)_{B-L}\to U(1)_Y$ at the
Planck scale with the subsequent breaking $SU(2)_L\times U(1)_Y\to U(1)_{QED}$
at the TeV scale.  After the Planck scale symmetry breaking occurs, a
global $SU(2)_L\times SU(2)_R$ symmetry remains in the brane picture;
this breaks on the TeV--brane to a diagonal group $SU(2)_D$ 
corresponding to the custodial $SU(2)$ symmetry present in the SM
\cite{sundrum}.  

It has been shown \cite{warped,NomuraI,sundrum} that due to the presence 
of the $SU(2)_D$ custodial symmetry, this Warped Higgsless
Model (WHM) enjoys good agreement with precision EW data at the level
of a few percent.  However, it has been
argued \cite{Bar,NomuraII} and demonstrated \cite{DHLR} that the
region of parameter space in the WHM that results in good
agreement with the EW data leads to perturbative unitarity
violation (PUV) in $W_L^+W_L^-$ scattering at energies of order $\sim 2-3$
TeV.  Furthermore, a scan of the parameter space of the WHM
shows that the scale of perturbative unitarity violation is never significantly
raised, even in those
regions where comparisons with the precision measurements are
anticipated to be quite poor \cite{DHLR}. 
To restore unitarity in gauge boson scattering, additional new
physics is required at or below the RS cutoff of the effective theory
on the TeV--brane.  Even though this does not by itself rule out
the model, it suggests that interactions in the gauge sector are
problematic above the TeV scale.

To address some of these issues, the authors of
Ref.~\cite{CsakiTeV} have examined the effects of including
IR(TeV)--brane terms for the $U(1)_{B-L}$ and the custodial $SU(2)_D$ 
gauge symmetries.  It is well-known that the introduction of brane
terms can alter the couplings and masses of the corresponding Kaluza-Klein 
(KK) tower states \cite{DHRbt}
and this would hence affect their contributions to the precision EW 
observables and to $W_L^+W_L^-$ scattering.  These authors concluded
that the addition of the
$U(1)_{B-L}$ brane term could lead to improved agreement 
with the EW data,  and, in addition, lowers the mass of the lightest 
KK state to $\sim$ 300 GeV.  Light KK states are generically expected
to help restore perturbative unitarity in high energy gauge boson
scattering, however the analysis of Ref. \cite{CsakiTeV} did not quantify
this point.  

In this paper, we also study the effects of the IR--brane kinetic terms
associated with both the $U(1)_{B-L}$ and $SU(2)_D$ symmetries; 
here, we pay particular attention to low energy perturbative unitarity
violation.  For the $U(1)_{B-L}$ boundary term, we
find that the scale of PUV in the model is {\it independent} of
the size of the brane term.    We also demonstrate
that increasing the ratio of the 5--$d$ couplings, 
$\kappa\equiv g_{5R}/g_{5L}$, improves the agreement with the tree-level
SM relations in the electroweak sector, but
lowers the scale at which perturbative unitarity is violated, 
similar to our previous results \cite{DHLR}.  In the case of
the $SU(2)_D$ kinetic term, we find that perturbative unitarity violation
in $W_L^+W_L^-$ scattering could be delayed to center of mass energies 
of order $\sim 6-7$ TeV.  However, agreement with the tree-level SM
relations is rather poor, 
with the disparity worsening
as the size of the $SU(2)_D$ brane term increases.  In addition,
we compare the predictions for the lowest lying gauge KK state
to the searches for new gauge bosons at the Tevatron
Run I and II and for contact interactions at LEP II 
and find that the collider bounds
restrict the potential size of the IR--brane kinetic terms.
However, these  collider bounds allow for the PUV scale to approach
$6-7$ TeV.

We describe our setup in the next section.  The EW and collider
constraints are discussed in section 3.  Perturbative unitarity 
in this model is the subject of section 4 and our concluding remarks
are given in section 5.

\section{The Model}

Here, we briefly discuss the modifications induced in our earlier 
analysis \cite{DHLR} 
due to the presence of the $U(1)_{B-L}$ and $SU(2)_D$ brane terms; these 
changes are quite straightforward.  We employ the notation 
introduced in our previous work.  
In what follows, when we consider the 
effects of the $U(1)_{B-L}$ kinetic term we also include the UV--brane
terms associated with the $SU(2)_L$ and $U(1)_Y$ symmetries in our analysis;
these UV kinetic terms were included in our earlier results.
However, for simplicity, we omit the UV terms in our study of the
$SU(2)_D$ kinetic term.

The introduction of new kinetic terms on the TeV brane leads to a 
shift in the original action (given in Eq.(4) of Ref.{\cite {DHLR}}) by
an amount 
\begin{equation}
\delta S_{brane}= \int d^4x dy \sqrt{-g} ~\delta (y-\pi r_c) 
\Big[-{1\over {4}} r_c c_B F_{B-L}^2 -{1\over 4(g_{5L}^2+g_{5R}^2)} r_c c_D 
(g_{5R}F_L+g_{5L}F_R)^2\Big]\,,
\end{equation}
with $g_{5L(R)}$ being the 5--$d$ $SU(2)_{L(R)}$ gauge coupling, $\pi r_c$ is
the brane separation in the RS model, and $c_{B,D}$ are dimensionless
parameters which quantify the size of the IR--brane kinetic terms.
Here $F_{B-L}$ is the field strength tensor for the 
$U(1)_{B-L}$ gauge field, and similarly $F_{L,R}$ corresponds to
$SU(2)_{L,R}$.  For later purposes it is convenient to 
introduce the quantities $\delta_{B,D}\equiv kr_c c_{B,D}/2$ as in our 
earlier analysis where $k$ is the RS curvature parameter. 
We next observe that a non-zero value for $\delta_B$ will alter the 
$\partial_z B=0$ boundary condition \cite{warped} on the TeV brane; 
instead, we now find $\partial_z B-\delta_B x_n^2 k\epsilon B=0$,
where $x_n$ represents the roots defining the KK spectra,
$\epsilon\equiv e^{-\pi kr_c}$, and $B$ represents the $U(1)_{B-L}$ 
gauge field. A similar shift is observed in the case of the 
combination of fields associated with the $SU(2)_D$ brane term, \ie, 
$\partial_z(g_{5R}A_L+g_{5L}A_R)-\delta_D x_n^2 k\epsilon (g_{5R}A_L+
g_{5L}A_R)=0$. 
Solving these new boundary conditions leads to alterations of the 
wavefunction coefficients as well as the eigenvalue equations for the 
KK tower masses. It is important to note, however, that the 
$W^\pm$ KK tower masses and couplings are left unaltered 
by a non-zero value of $\delta_B$, but are modified by $\delta_D$.
 
In calculating the couplings to both fermions and 
$W^\pm$ pairs for the photon, the $Z$, as well as the rest of the KK 
tower states,  
one of the dominant effects due to the new brane terms is the shift in 
the normalizations of the $W_n^\pm$ and $Z_n$ wavefunctions. These 
normalizations 
now pick up additional terms; for the case of the $Z_n$, in comparison to 
our earlier result
(Eq.(50) of Ref.{\cite {DHLR}}); we now obtain: 
\begin{eqnarray}
N_{Z_n} & = & \int_R^{R'}dz\, {R\over z}\, \left\{
|\chi_L^n(z)|^2(2+c_Lr_c\delta(z-R))+2|\chi_R^n(z)|^2
+2|\chi^n_B(z)|^2\right.\nonumber\\
& + & \left. c_Yr_c{|\kappa\chi^n_B(z)+\lambda\chi^n_R(z)
|^2\over\kappa^2+\lambda^2}\delta(z-R)+c_Br_c |\chi_B^n(z)|^2
\delta(z-R')/\epsilon \right.\nonumber\\
& + & \left. c_D r_c{|\kappa \chi^n_L(z)+\chi^n_R(z)|^2\over {1+\kappa^2}}
\delta(z-R')/\epsilon\right\}\,,
\end{eqnarray}
where $\chi^n_i$ are the wavefunctions for the relevant gauge KK state,
and $\lambda$ is defined as the ratio $\lambda\equiv g_{5B}/g_{5L}$.
A similar shift in the $W_n^\pm$ normalization also occurs,
\begin{eqnarray}
N_{W_n} & = & \int_R^{R'}dz\, {R\over z}\, \left\{
|\chi_L^n(z)|^2(2+c_Lr_c\delta(z-R))+2|\chi_R^n(z)|^2
\right.\nonumber\\
& + & \left. c_D r_c{|\kappa \chi^n_L(z)+\chi^n_R(z)|^2\over {1+\kappa^2}}
\delta(z-R')/\epsilon\right\}\,.
\end{eqnarray}
These new TeV brane terms also lead to additional contributions to the 
normalization of the photon wave function,
\begin{equation}
N_\gamma=2\pi r_c\alpha_L^2\left( {\kappa^2+\lambda^2+\kappa^2
\lambda^2\over\kappa^2\lambda^2}\right)\left\{ 1+{1\over\pi kr_c}
{\kappa^2\lambda^2\delta_L+\kappa^2 \delta_B+(1+\kappa^2)\lambda^2 \delta_D
+(\kappa^2+\lambda^2)\delta_Y\over
\kappa^2+\lambda^2+\kappa^2\lambda^2}\right\}\,,
\end{equation}
where $\alpha_L$ is a numerical constant which is determined from the
boundary conditions and appears in the KK decomposition of the
$A_L^3$ gauge field.
Due to the abelian nature of the $U(1)_{B-L}$ group,  new brane term 
contributions to the $W$ 4-point or gauge 3-point functions do not occur.  
However, such contributions are induced in the case of the 
$SU(2)_D$ brane term. 

\section{Precision Measurements and Collider Bounds}

Our analysis now proceeds by analogy with our earlier work \cite{DHLR}: 
we hold $M_{W,Z}$ as well as the UV--brane kinetic terms 
$\delta_{L,Y}$ fixed and explore the parameter space spanned by 
the parameters $\kappa$ and $\delta_{B,D}$. 

\begin{figure}[htbp]
\centerline{
\includegraphics[width=8cm,angle=90]{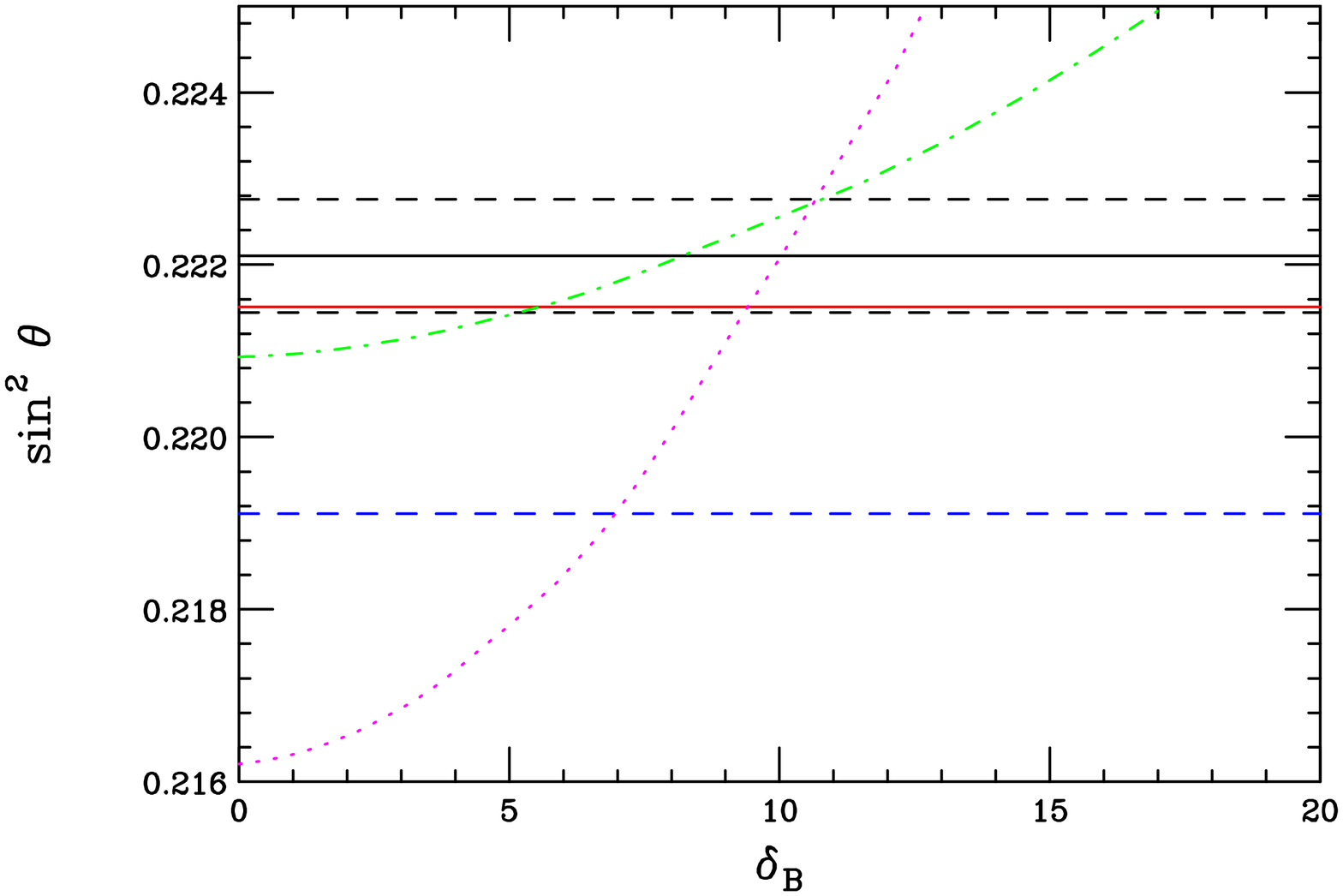}}
\vspace*{0.45cm}
\centerline{
\includegraphics[width=8cm,angle=90]{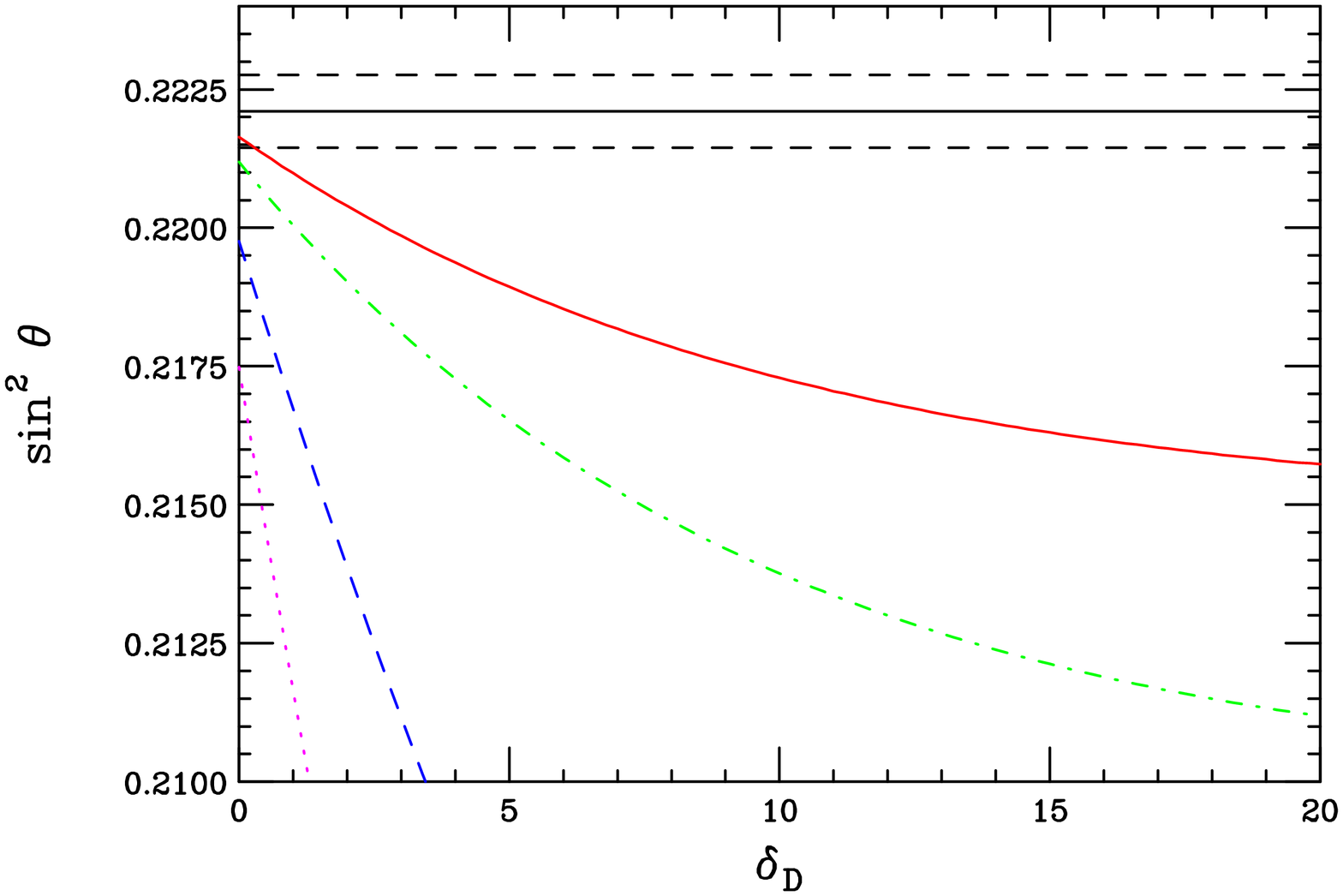}}
\vspace*{0.25cm}
\caption{$\sin^2 \theta$ in each of the three definitions as a function 
of $\delta_{B,D}$.  The black horizontal solid and 
dashed curves correspond to the on-shell value $\pm 1\sigma$,
the solid red (dashed blue) curve represents $\sin^2 \theta_{eff}$ for 
$\kappa=3 (1)$ while the dash-dotted green (dotted magenta) curve 
is for $\sin^2 \theta_{eg}$.  The top (bottom) panel illustrates the
effects of including the $U(1)_{B-L}$ ($SU(2)_D$) kinetic term.  We take
only one IR kinetic term to be non-vanishing at a time.}
\label{sin2}
\end{figure}

We previously introduced the three different quantities related to 
the weak mixing angle: $\sin^2\theta_{os}=1-M_W^2/M_Z^2$,  
$\sin^2\theta_{eg}=e^2/g_{W_1}^2$  (where $g_{W_1}$ is the coupling of the 
particle we identify as the $W$ to the SM fermions), and 
$\sin^2\theta_{eff}$, which is defined at the $Z$-pole. All three must 
take on the same value at the tree-level in the SM. 
They can differ significantly 
in the present scenario; however, there are  preferred 
parameter space regions, \ie, when $\kappa$ is large \cite{DHLR},
that yield consistent values. 
The first question to address here  
is the variation of $\sin^2\theta_{eg,eff}$ with respect to the fixed on-shell 
value, $\sin^2 \theta_{os}$, as $\delta_{B,D}$ are allowed to change 
for fixed $\kappa$. The results of 
this analysis are shown in Fig.~\ref{sin2}. In the top panel we observe 
that $\sin^2\theta_{eff}$ is $\delta_B$-independent, which we have
verified analytically, while 
$\sin^2\theta_{eg}$ increases as $\delta_B$ increases. In fact we see that 
for $\kappa=1(3)$ excellent agreement between the on-shell and effective 
values is obtained when $\delta_B \simeq 8(10)$. Overall, however, the case 
$\kappa=3$ yields more consistent values, as in our earlier work, 
due to the large separation between the quantities $\sin^2\theta_{os}$ 
and $\sin^2\theta_{eff}$  
when $\kappa=1$. Clearly, a non-vanishing value of $\delta_B$ does help 
to bring the values of the various definitions of  $\sin^2\theta$ into
agreement. In contrast, in the bottom panel we see that as $\delta_D$ 
increases both of the different $\sin^2 \theta$ values shrink in size and 
move away from the on-shell value thus getting further from the SM limit. 
Of course the $\kappa=3$ values remain closer to the SM than do those for 
$\kappa=1$, but in all cases the agreement is poor.
%
\nn

\begin{figure}[htbp]
\centerline{
\includegraphics[width=8cm,angle=90]{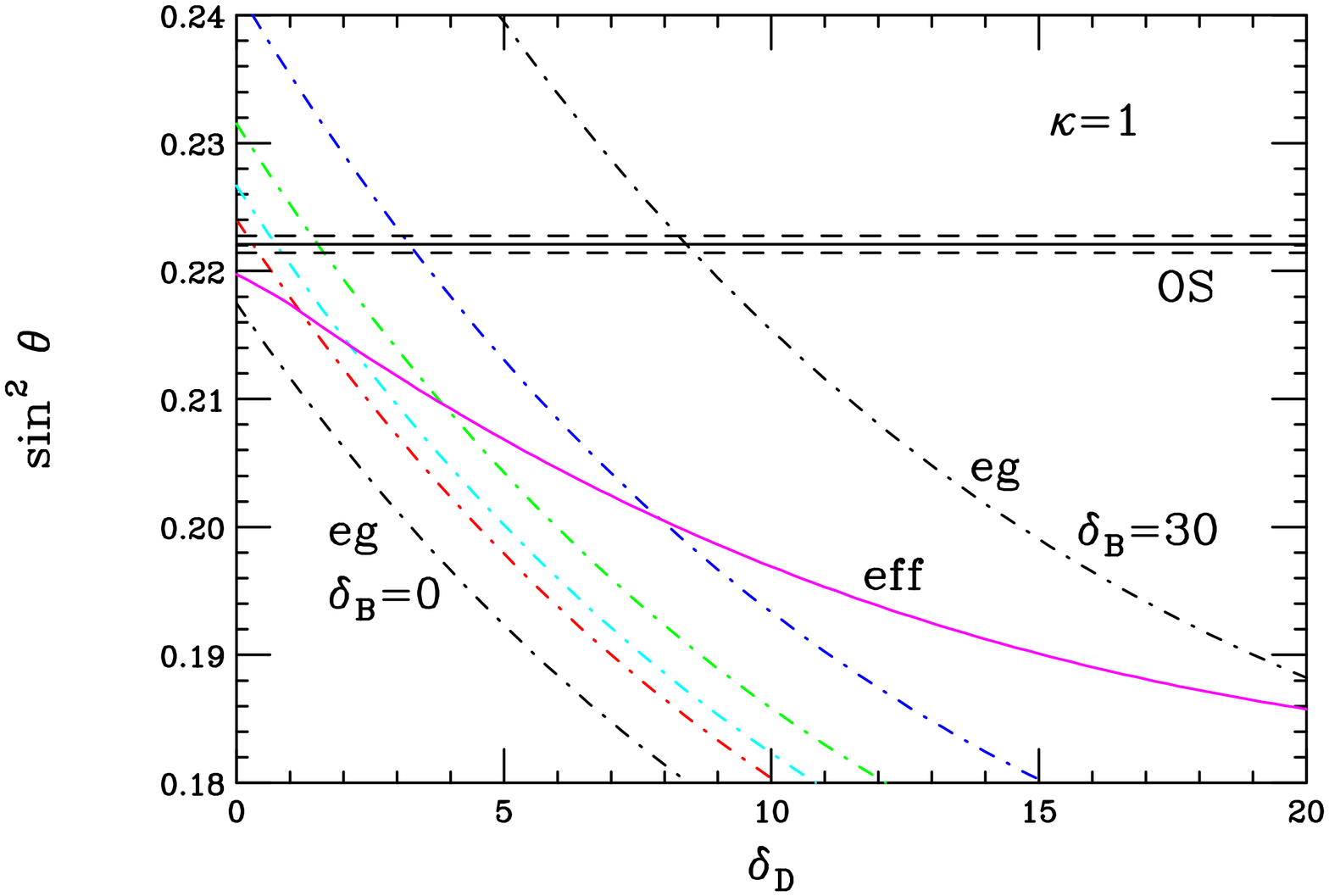}}
\vspace*{0.25cm}
\caption{Same as in the previous figure but now with both $\delta_{B,D}$ 
nonzero for the case $\kappa=1$. The solid magenta curve is the value of 
$\sin^2 \theta_{eff}$ while the dash-dotted curves are all for 
$\sin^2 \theta_{eg}$ for, from left to right, $\delta_B=0,10,12,15,20$ and 
30, respectively.}
\label{sin2both}
\end{figure}

Since $\delta_{B,D}$ shift the the $\sin^2 \theta_{eg}$ curves in opposite 
directions, it is interesting to see what happens when these brane terms are 
simultaneously nonzero. This can be seen in Fig. \ref{sin2both} for
$\kappa=1$. For the 
range of $\delta_D$ of interest we see that we can always find a value of 
$\delta_B$ for which $\sin^2 \theta_{os}\simeq \sin^2 \theta_{eg}$. 
Unfortunately, since $\sin^2 \theta_{eff}$ is $\delta_B$ independent, 
including this brane term does not bring this quantity into accord with
the others for $\kappa=1$;  larger values of $\kappa$ may help in this
regard.
\nn

\begin{figure}[htbp]
\centerline{
\includegraphics[width=8cm,angle=90]{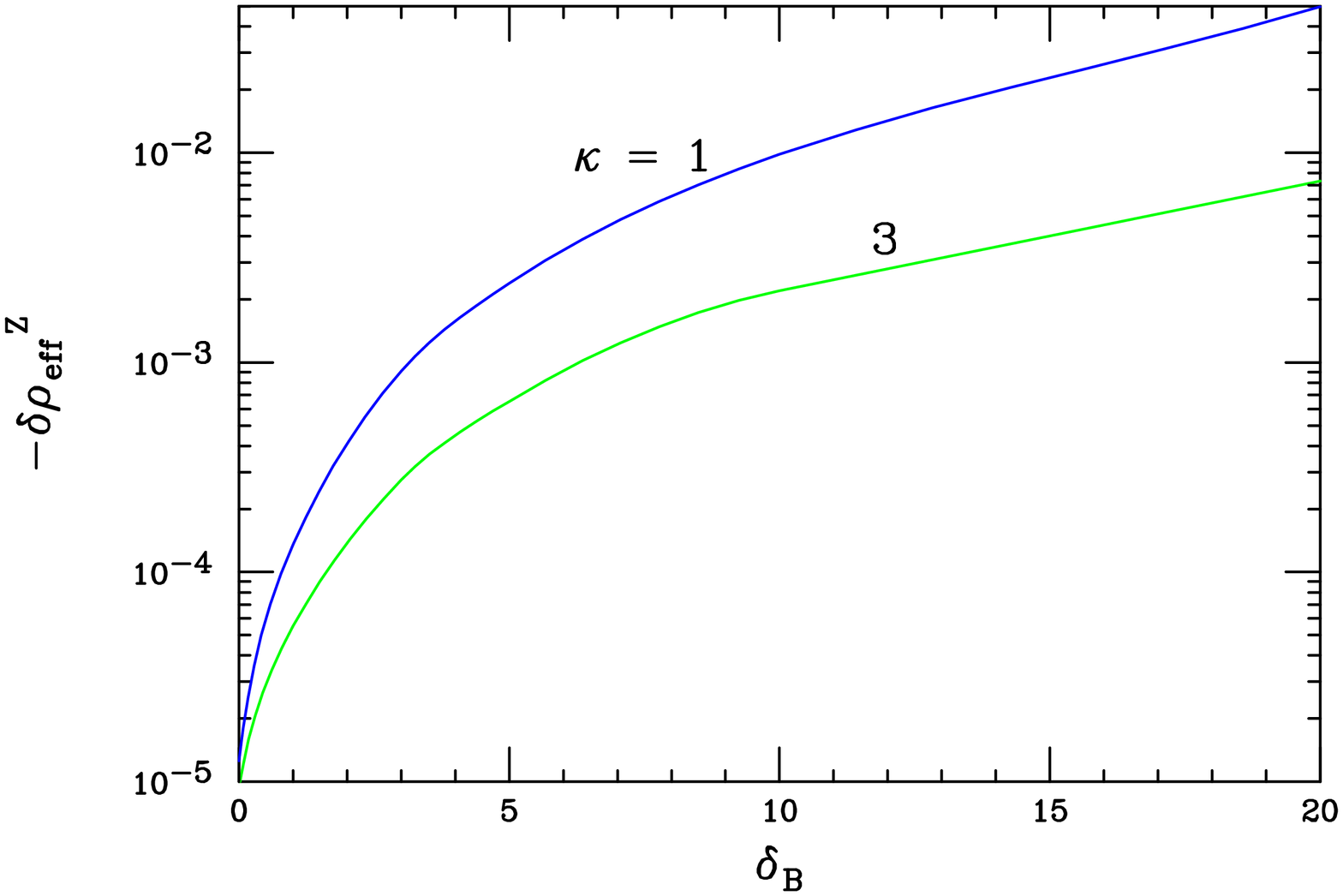}}
\vspace*{0.45cm}
\centerline{
\includegraphics[width=8cm,angle=90]{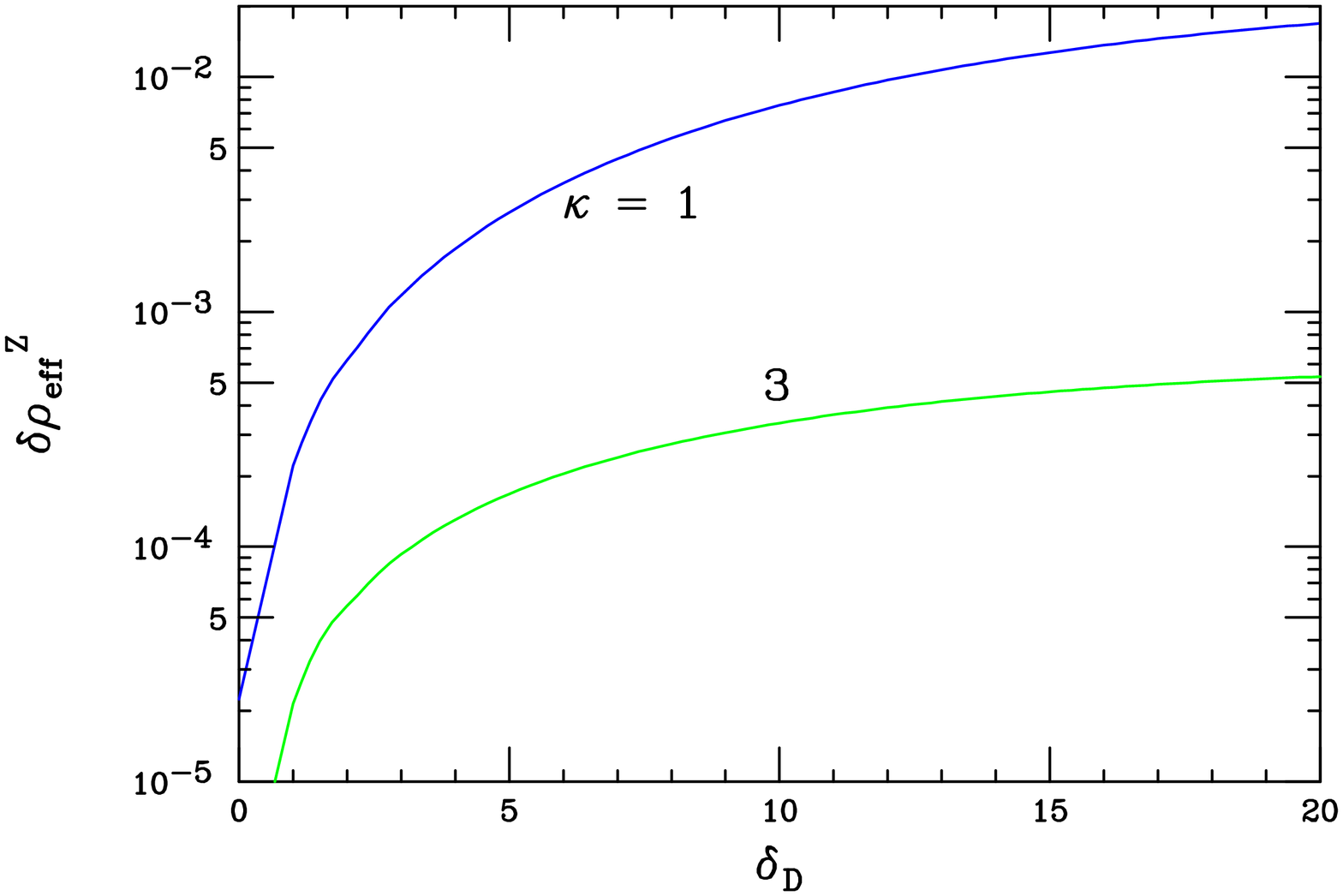}}
\vspace*{0.25cm}
\caption{ $\delta \rho_{eff}^Z$ as a function of $\delta_{B,D}$ for 
$\kappa=1$ and 3.   We take
only one IR kinetic term to be non-vanishing at a time.}
\label{drho}
\end{figure}

Another quantity of interest is the value of the overall strength of the 
$Z$ boson coupling, denoted as $\rho_{eff}^Z$, and in particular, 
its deviation from unity, \ie, $\delta \rho_{eff}^Z \equiv \rho_{eff}^Z-1$.
This deviation is related to the pseudo-oblique parameter $T^*$ as   
$T^*/\alpha\equiv\delta \rho_{eff}^Z$. The pseudo-oblique parameters 
are defined in such a way so that they all take on the value zero in the 
tree level SM.  They are introduced to guide our thinking about
the direction in parameter space which approaches the SM.
We note that it is important {\it not} to confuse these 
pseudo-oblique parameters with the conventionally defined purely
oblique $S,T,U$.  The dependency of  $\delta \rho_{eff}^Z$  on 
$\delta_{B,D}$ for two different values of $\kappa$ is shown in 
Fig. \ref{drho}. Note that this parameter remains relatively small 
in magnitude for both values of $\kappa$ as long 
as either $\delta_{B,D}$ does not become too large.  

\begin{figure}[htbp]
\centerline{
\includegraphics[width=8cm,angle=90]{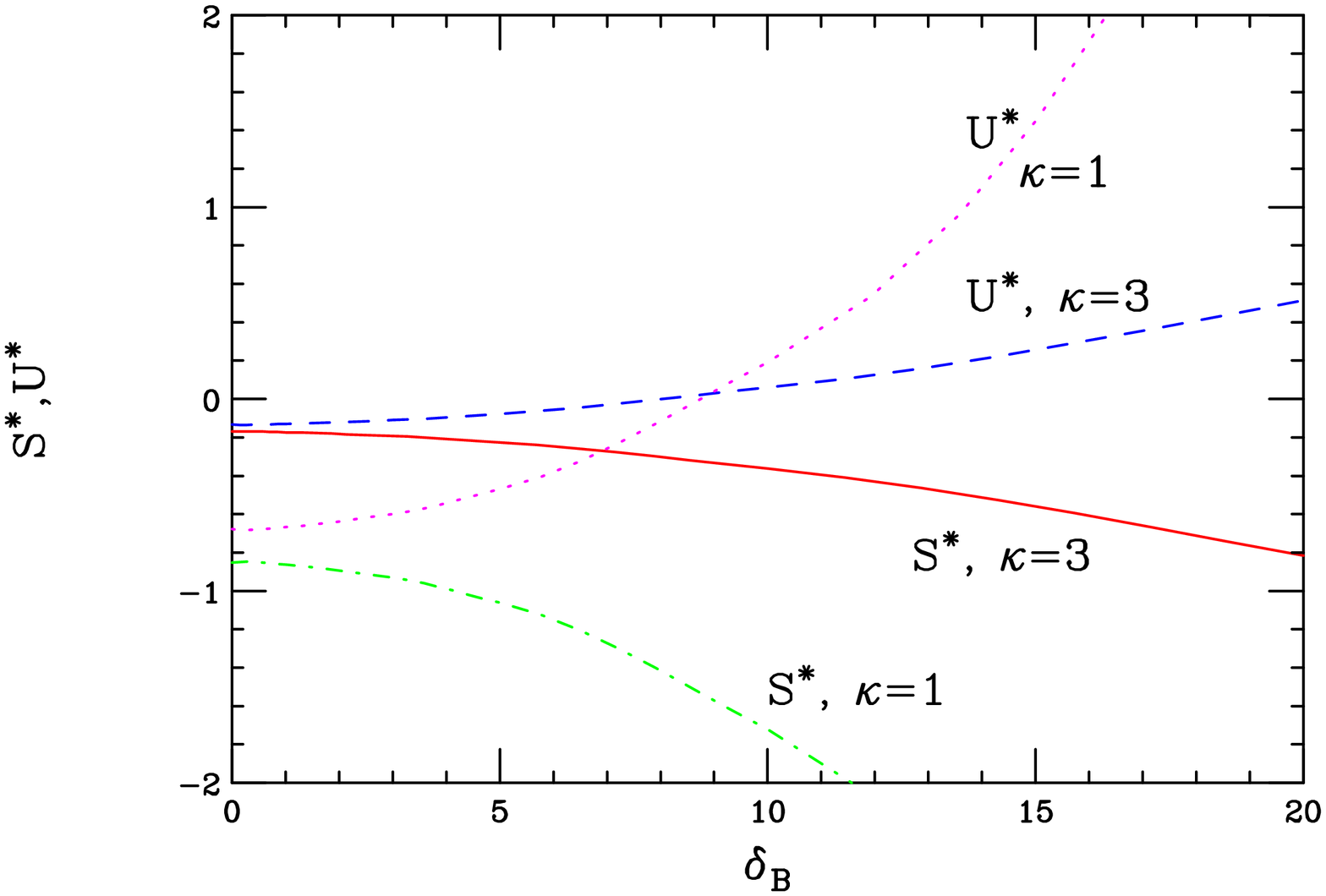}}
\vspace*{0.45cm}
\centerline{
\includegraphics[width=11.7cm,angle=0]{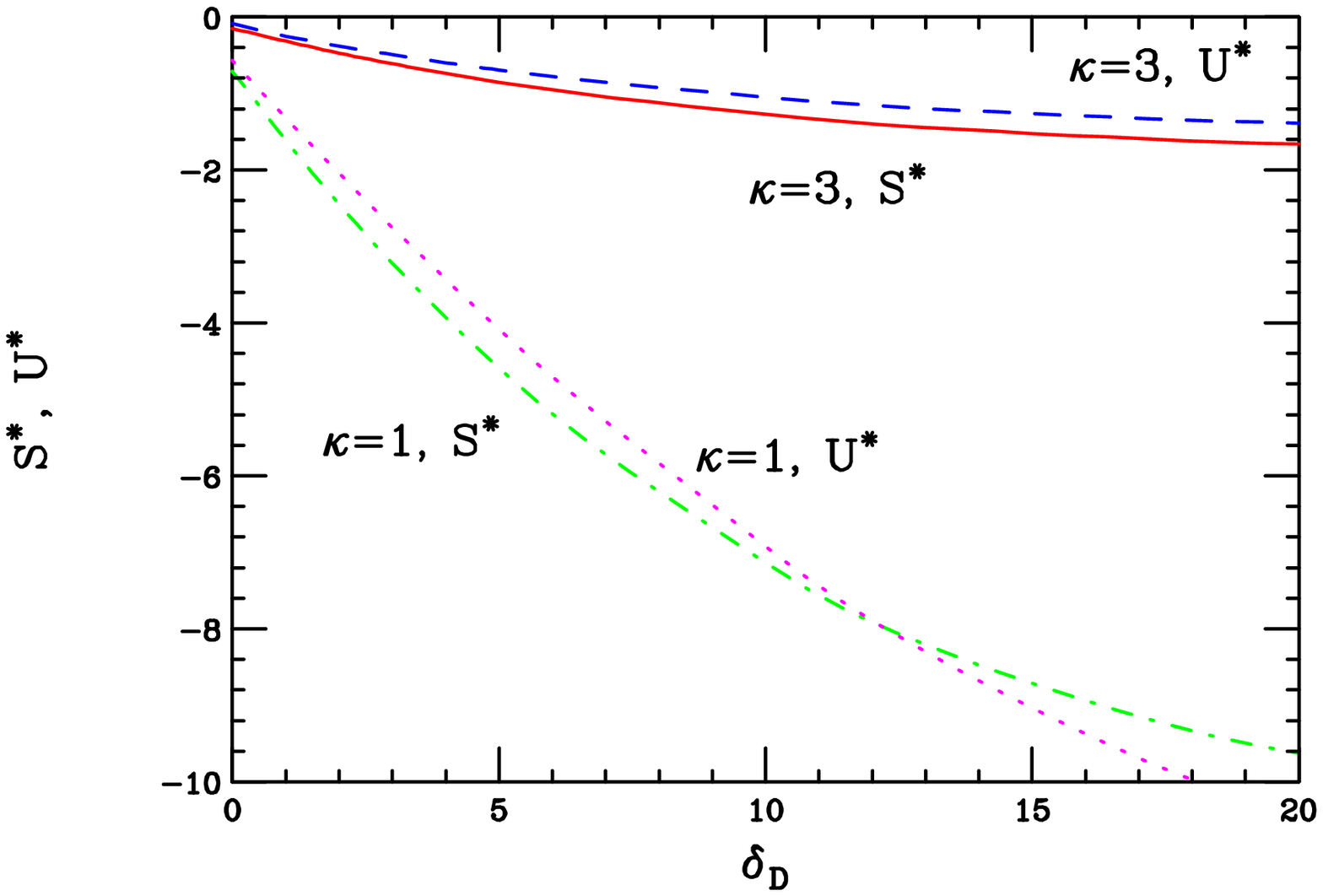}}
\vspace*{0.25cm}
\caption{Values of the pseudo-oblique parameters $S^*$ (solid red,
dash dotted green) and $U^*$ (blue dashed, dotted magenta) for  
of $\kappa=(3,1)$ as labeled as functions of $\delta_{B_D}$.  We take
only one IR kinetic term to be non-vanishing at a time.}
\label{pstu}
\end{figure}

The other pseudo-oblique parameters
$S^*,U^*$, as defined in Ref. \cite{DHLR}, are also functions of 
$\delta_{B,D}$ as shown in  
Fig. \ref{pstu}. For $U^*$, some values of $\delta_B$ 
improve the agreement with the SM limit, 
while $S^*$ tends away from its SM value.
We see that, overall, smaller values of $\delta_B$ are 
again preferred. In the case of $\delta_D$ we see that both $S^*$ and $U^*$ 
move away from the SM limit with the shifts being much more significant 
in the case of $\kappa=1$. 

Our approach to calculating the pseudo-oblique observables, $S^*T^*U^*$, 
differs from that of $STU$ as calculated by Csaki \etal {\cite {CsakiTeV}}. 
In our approach, we numerically fix the masses of the first charged and 
neutral gauge KK excitations to be those of the physical $W$ and $Z$ bosons 
observed at colliders.  We use these as input to our analysis,
together with the strength of the charged 
current coupling determined by $G_F$. From these the couplings 
and masses of all the gauge KK states can be obtained. 
The pseudo-oblique parameters are then defined 
in terms of observables via the $W$ mass, the invisible width of the $Z$ and 
the fermionic couplings determined at the $Z$-pole. $S^*T^*U^*$ 
are chosen to 
vanish at the tree-level in the SM. Csaki \etal\ choose a different 
scheme wherein the SM gauge couplings $g$ and $g'$ are used as 
input parameters together with the usual relationship $1/e^{2}=
1/g^{2}+1/g'^{2}$.  This fixes $\sin^2 \theta$ and hence 
the couplings of the $W$ and $Z$. 
From this the $W$ and $Z$ and other KK masses, as well as their
couplings, can be determined. The $STU$ parameters in Ref. \cite{CsakiTeV} 
can then be calculated as shifts in the masses as well as the 
wavefunctions and normalizations for
the $W$ and $Z$. It is clear that these two sets of electroweak 
parameters probe different relationships between the masses and couplings of 
the $W$ and $Z$ described by distinct choices of input parameters. In either 
case they allow for a measure of how far the model predictions are from the 
tree level SM. However, without employing the full loop corrections and 
overcoming the problem of `subtracting out' the Higgs loop effects
(described in \cite{DHLR}) neither 
set of parameters can be directly compared with data. 
\nn

%
\nn

\begin{figure}[htbp]
\centerline{
\includegraphics[width=11.7cm,angle=0]{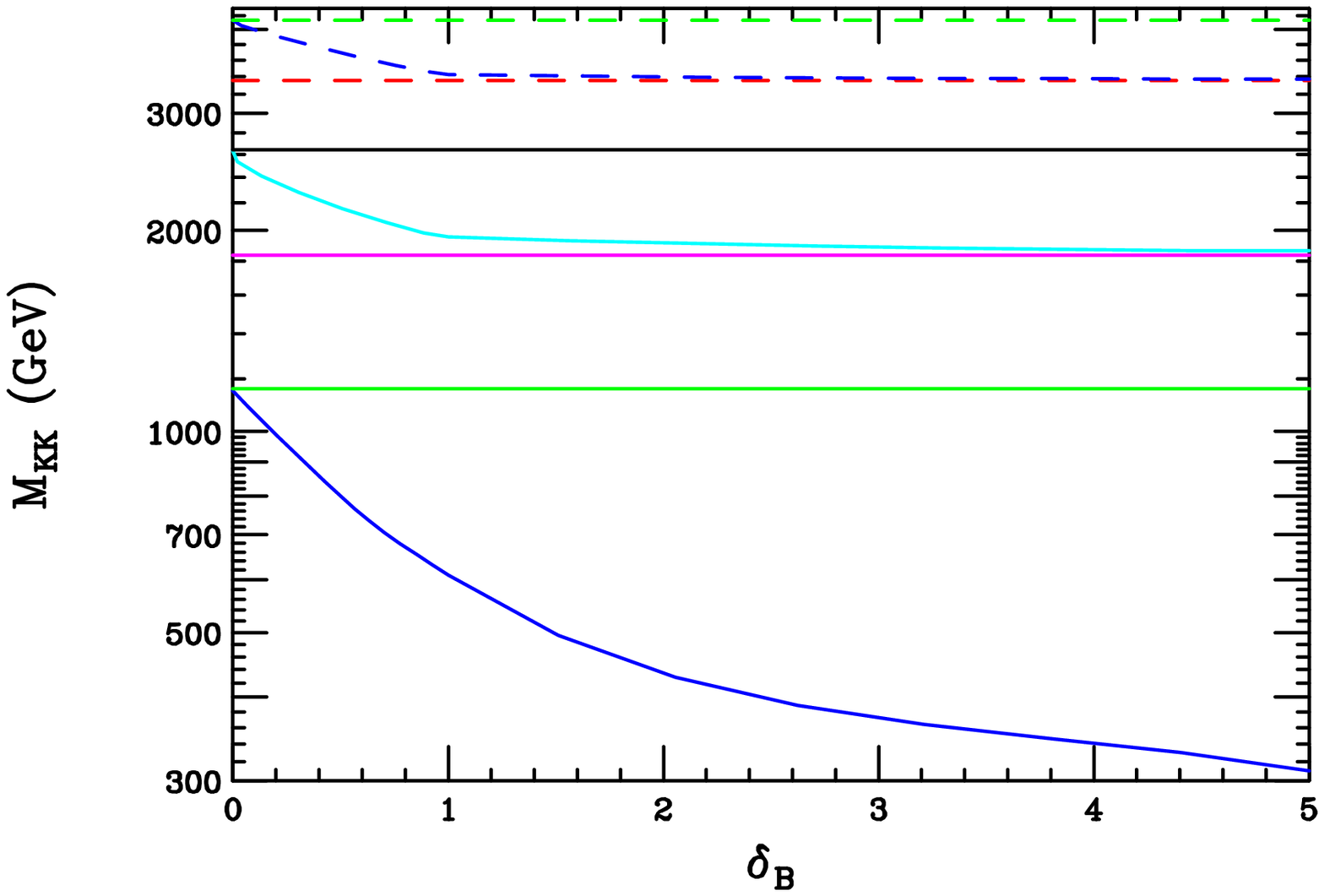}}
\vspace*{0.45cm}
\centerline{
\includegraphics[width=11.7cm,angle=0]{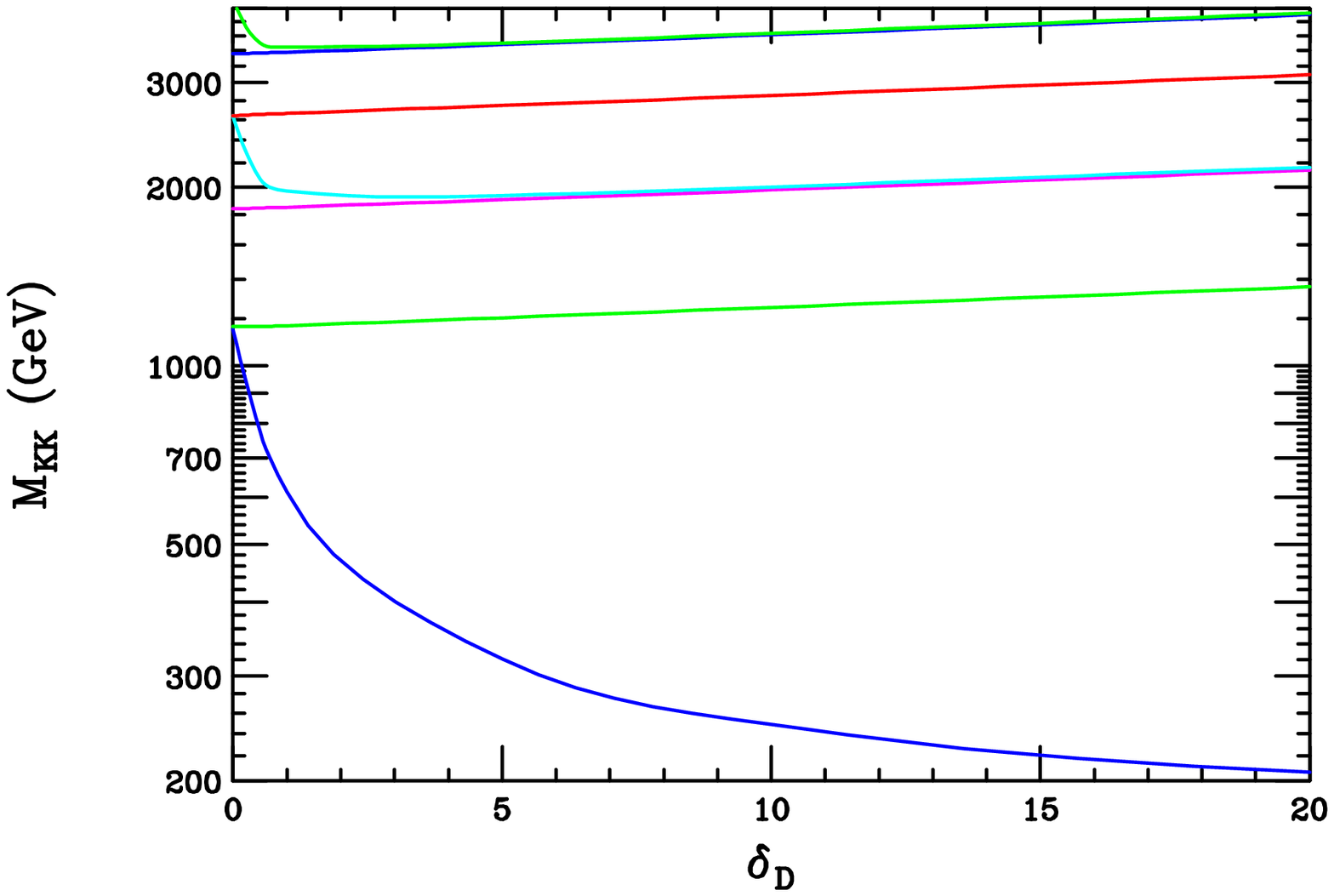}}
\vspace*{0.25cm}
\caption{ Behavior of the neutral KK mass spectrum as a function of 
$\delta_{B,D}$. From bottom to top on the left the curves correspond to the 
states $Z_{2,3,..}$. $\kappa=1$ has been assumed.  We take
only one IR kinetic term to be non-vanishing at a time.}
\label{kkspec}
\end{figure}

To go further in the analysis of this model, we need
to consider how non-zero values of 
$\delta_{B,D}$ lead to modifications of the KK spectra. 
Clearly, the $U(1)_{B-L}$ 
brane term does not influence the $W$ KK tower so we turn our 
attention to the neutral KK states.  
The major effect of a non-zero $\delta_B$ on neutral KK states  
can be clearly seen in the upper panel 
of Fig. \ref{kkspec} for the case of $\kappa=1$; 
the same qualitative behavior 
occurs for other values of $\kappa$.  
Here we immediately observe that the single, non-degenerate states are 
unaffected while one member of the nearly degenerate paired states, 
the one which couples mainly to $B-L$,    
gets its mass reduced as $\delta_B$ is increased. The remaining member of
the pair stays unaffected. In particular, we see that the state 
$Z_2$ becomes light (note that here, $Z_1$ is the lightest state and
corresponds to the SM $Z$). 
Further increasing $\delta_B$ leads to the appearance of new sets 
of almost degenerate pairs of states. Including $\delta_D$ has the opposite 
effect in that the member of the pair which couples mainly to $T_{3L}$ 
gets its mass lowered. The other states are only slightly affected.  
In the case of $SU(2)_D$, the charged KK states have 
all of their masses lowered in analogy with the falling curves in the lower 
panel\footnote{Note that the root corresponding to the 
observed $W$ is also lowered.  Since the mass scale of the heavier
KK states is obtained by matching this root to $M_W$,  the mass of the
neutral states rises.}.
This figure demonstrates
that the $U(1)_{B-L}$ and $SU(2)_D$ brane terms are at least partly doing 
what we had expected, \ie, lowering 
the KK masses so that the now lighter states can 
have a potentially greater influence on unitarity in $W_L^+W_L^-$ scattering. 
They do, however, lower the masses of {\it different} sets of KK states and 
this is critical for unitarity considerations as we will see below.  
\nn

\begin{figure}[htbp]
\centerline{
\includegraphics[width=8cm,angle=90]{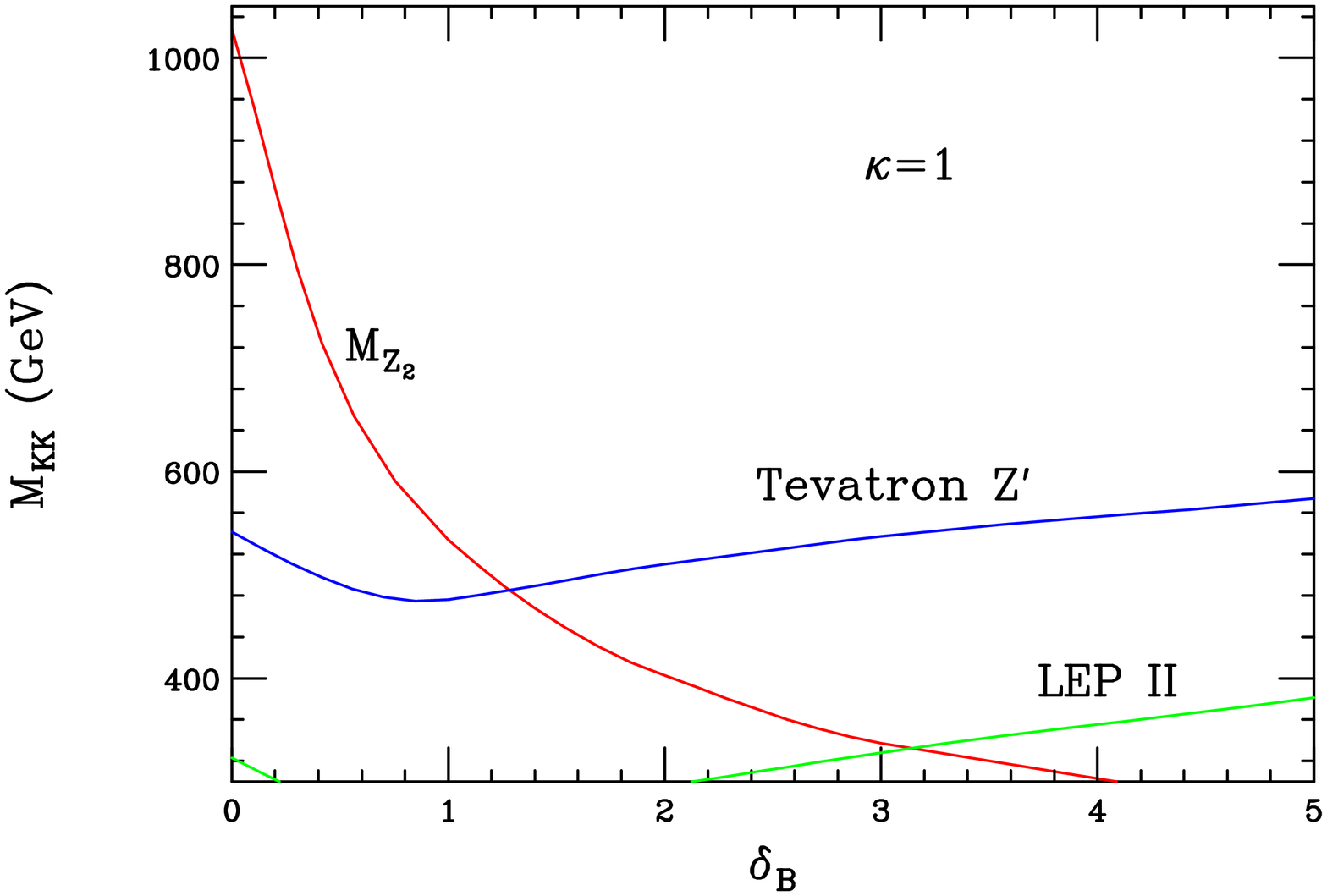}}
\vspace*{0.45cm}
\centerline{
\includegraphics[width=8cm,angle=90]{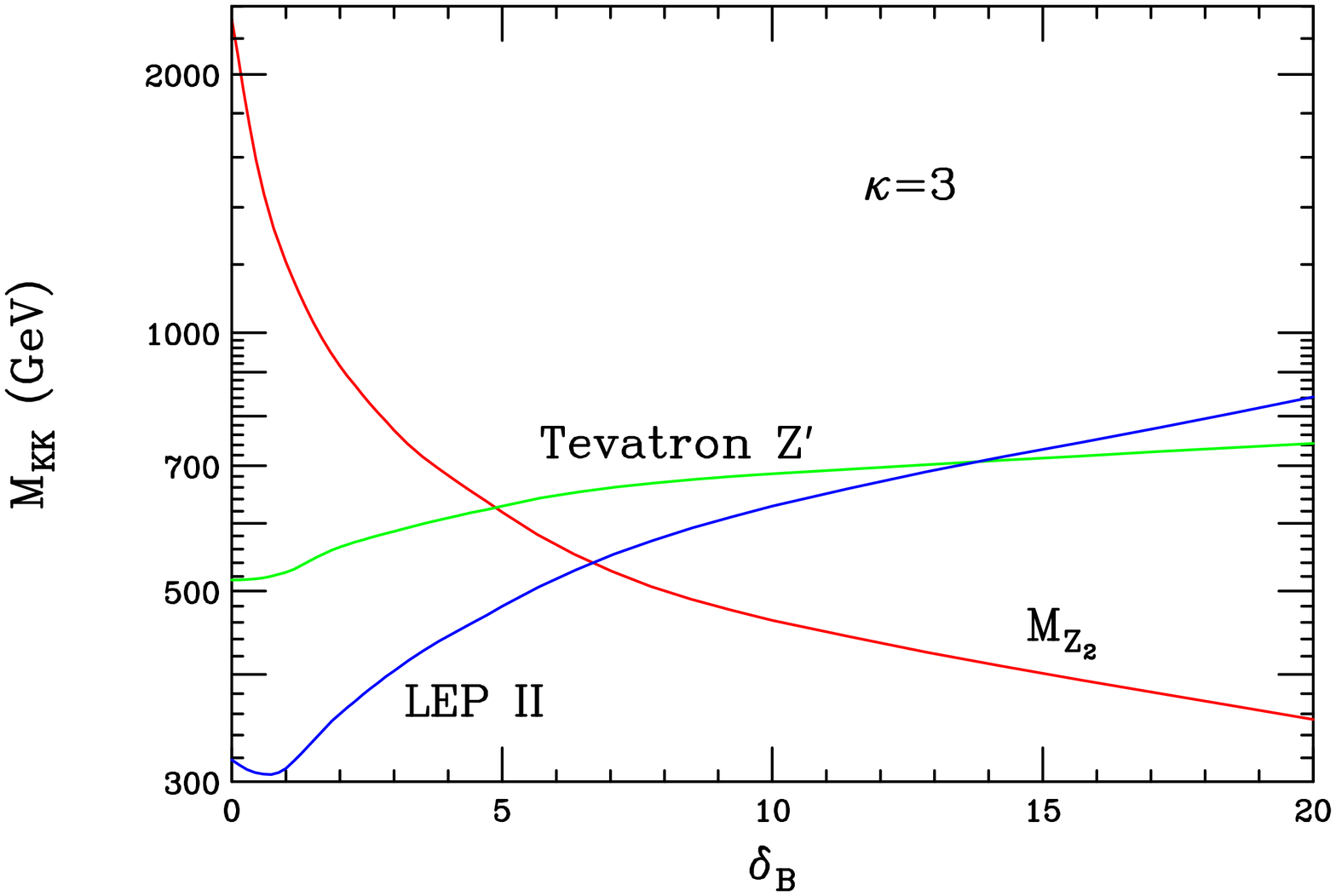}}
\vspace*{0.25cm}
\caption{The predicted mass of the lightest KK excitation, the lower 
bound on the mass from the Run II Tevatron $Z'$ searches as well as the lower 
bound from LEPII as a function of $\delta_{B}$, assuming $\delta_D=0$. 
The collider limits are discussed in detail in the text.}
\label{collbl}
\end{figure}

\begin{figure}[htbp]
\centerline{
\includegraphics[width=8cm,angle=90]{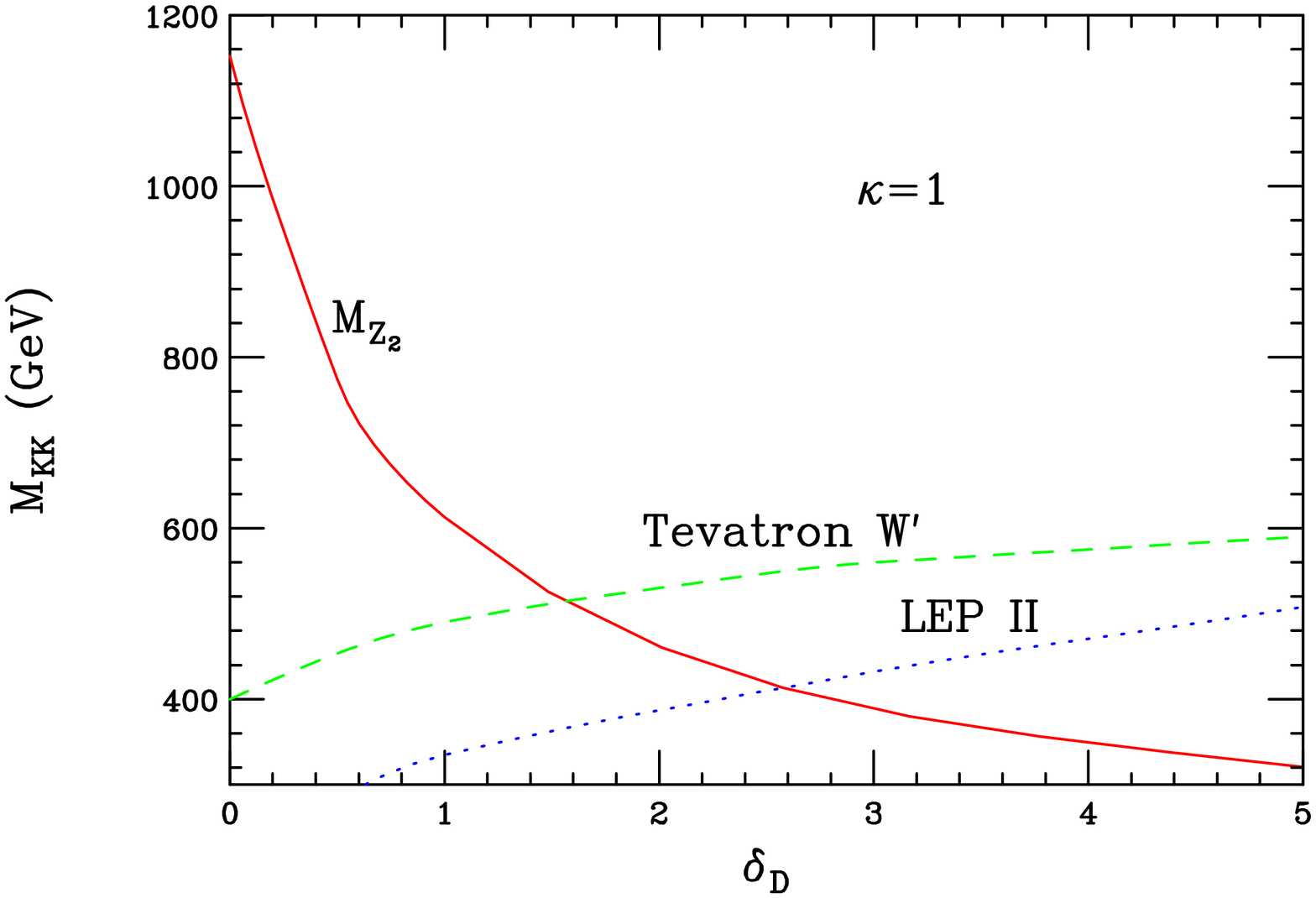}}
\vspace*{0.45cm}
\centerline{
\includegraphics[width=8cm,angle=90]{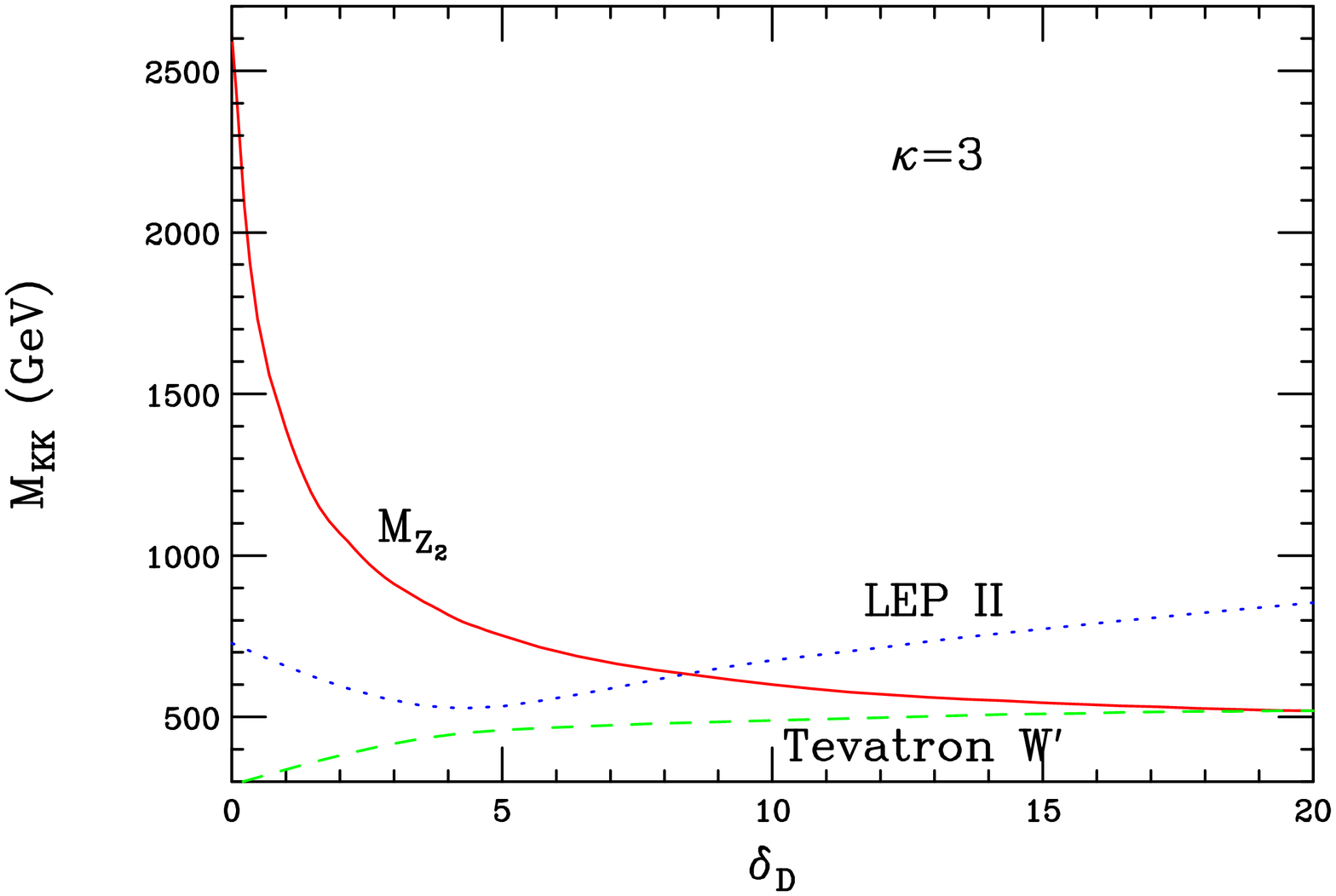}}
\vspace*{0.25cm}
\caption{The predicted mass of the lightest KK excitation, the lower 
bound on the mass from the Run I Tevatron $W'$ searches as well as the lower 
bound from LEPII as a function of $\delta_D$, assuming $\delta_B=0$.
The collider limits are discussed in detail in the text.}
\label{colldl}
\end{figure}

One may wonder, since some of the KK states are becoming so light, if there 
are conflicts with direct searches for new vector bosons 
at the Tevatron as well as with
indirect searches such as those for contact interactions at, \eg, LEP II. 
We recall that while the Tevatron experiments search for new 
resonances decaying into leptons via Drell-Yan production, 
the LEP bounds result from searches for deviations in cross sections and 
angular distributions from SM expectations below production threshold. 
In the case of the charged KK states, whose masses are lowered by the 
$SU(2)_D$ brane terms, the best limit comes from the Run I 
search at the Tevatron {\cite {oldtev}}. 
The strongest bounds on the direct production of $Z'$-like states come 
from Run II data using $200~$pb$^{-1}$ 
of integrated luminosity {\cite {TeV2}}, while indirect bounds on such states 
have been supplied by the LEPEWWG {\cite {LEP2}}. All of these sets of data  
have been employed in obtaining the results which follow. Figures~\ref{collbl} 
and~\ref{colldl} 
show the $\delta_{B,D}$ dependence of the lightest KK excitation 
mass for $\kappa=1,3$ as well as 
the corresponding bounds on this state from LEP II and the Tevatron. 
The non-trivial nature of these bounds arises from the modification in 
the $W_2$ and 
$Z_2$ couplings as $\delta_{B,D}$ are varied.  Note that in the case of
an $SU(2)_D$ [$U(1)_{B-L}$] brane term, the best limit from the Tevatron
arises from constraints on $W'$ [$Z'$] production.  The reason for this
is that, in the case of $SU(2)_D$, both $W$ and $Z$ KK excitations may
be light and the Tevatron constraints on $W'$ production are generally
stronger than those for $Z'$ production since the
cross section times leptonic branching fraction is larger in the $W'$
case.  For both values of $\kappa$ we 
again see that smaller values of $\delta_{B,D}$ are in better agreement
with the data. Note that while 
the Tevatron bounds are somewhat sensitive to the assumption that all the SM 
fermions are localized close to the Planck brane due to possible variations
in the width of the $W_2$ and $Z_2$, this is {\it not} true for 
those from LEP. For example, one can imagine that for model building 
purposes, the right-handed top-quark might be moved away from the 
Planck brane; this could significantly 
alter the bounds from the Tevatron but those from LEP II would remain intact. 

As we will see below, the masses of the first $W$ and $Z$ KK excitations
must be relatively light, $\leq 1$ TeV, for there to be any impact on PUV.
Though their couplings to the SM fermions are somewhat reduced, such states
will not escape detection at the LHC and may even be observed in the near
future at the Tevatron. The first neutral KK state may be sufficiently 
light to be produced on resonance at a TeV-scale linear collider.
%
\nn

\nn

\section{Unitarity in $W_L^+W_L^-$ Scattering}

\begin{figure}[htbp]
\centerline{
\includegraphics[width=11.7cm,angle=0]{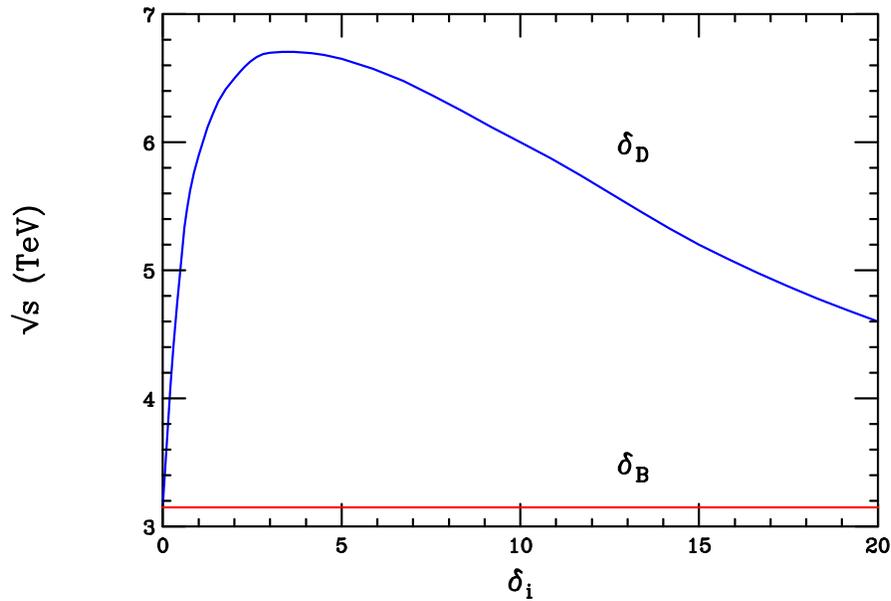}}
\vspace*{0.25cm}
\caption{The scattering energy at which perturbative unitarity is violated
in $W_L^+W_L^-$ scattering as a function of the kinetic terms.  We
take $\kappa=1$.}
\label{puv}
\end{figure}

We will now investigate the question of whether perturbative unitarity 
is preserved in this model.
As before, we examine the amplitude for the $W_L^+ W_L^- \to W_L^+ W_L^-$ 
elastic scattering process. In Ref.~{\cite {flat}},  
two sum rules were derived that 
insure the cancellation of terms growing with energy at high energy. Here,
we find that, as in our previous analysis, these sum 
rules are satisfied to good precision once sufficient KK states are included. 
However, these sum rules are technically only valid at infinite center of
mass energy. If the scattering occurs at a finite value of $\sqrt{s}$, then 
the amplitude cannot receive contributions from states much heavier than
$\sqrt{s}$. Therefore, we investigate the full amplitude in detail 
in the intermediate energy region, between $m_Z$ and the 
high-energy regime where the sum rules are valid. If unitarity is violated 
it will be in this region. Since the relevant expansion parameters, 
$M_{KK}^2/s$, are not small, we use the full tree-level 
amplitude from Ref. {\cite {Duncan}}. We numerically calculate the 
couplings using the $\delta_{B,D}$ generalized versions of Eq. (67) 
from Ref.~{\cite {DHLR}}. We then 
numerically evaluate the amplitude and apply the 
partial wave unitarity condition $|\Re a_0| \le 1/2$, 
where $a_0$ is the zeroth partial wave amplitude. The couplings were 
obtained independently on two different computing platforms, 
Maple and Mathematica. The partial wave amplitude was computed 
independently by three calculations, using Mathematica and Fortran.
We have included all KK states with masses up to $10 \tev$ and checked that the
results are stable against including more states. 
For $\delta_B\neq 0\,, \delta_D=0$ we 
find that perturbative unitarity is violated  
and, furthermore, the scale of PUV 
is {\it independent} of $\delta_B$. For $\kappa =1$ the violation 
occurs at $3.8\ \tev$; for $\kappa =3$ it occurs at $1.9\ \tev$, close
to the SM value.
We have also checked the case $\delta_L=0,\delta_B=4,\kappa=1$ which roughly 
corresponds to the case studied in {\cite {CsakiTeV}}; we found
PUV at 3.15 TeV.  For non-zero $\delta_D$, with all other $\delta_i$ set
to zero, we find that the scale of PUV is increased over some of
the parameter space, reaching energies $\sim 7$ TeV, 
as displayed in Fig. \ref{puv}.

These results can be understood heuristically. Naively, one expects that the
unitarity violations will be softened as the masses of the KK states
contributing to unitarity restoration are lowered. Hence, one expects that
a high value of $\delta_B$ will at least raise the scale of unitarity
violation. However, note that gauge boson scattering is a
fundamentally non-Abelian process. In the present model,  it is therefore
an $SU(2)_L$ process, and should not depend on the $U(1)_{B-L}$ dynamics. When
$\delta_B$ is turned on, the mass of one state in a pair responds
dramatically, while the other is unaffected. It is clear that the state
that responds should be predominantly a hypercharge boson, with very little
mixture of $W^3$ in its wavefunction. 
Indeed, we can write \cite{DHLR} the couplings of the neutral KK states 
to SM fermions as $(g_{Z_n}/c_w) (T_{3L}^f-s_n^2 Q^f)$. 
Calculation of the 
$s_n^2$ parameters confirms that the light state couples as a hypercharge 
boson. Numerically, we can look at 
the coupling of the light state to two $W_L$ bosons.
At $\delta_B=0$ this coupling is a factor of $6$ smaller than that
for the next neutral KK state, which is predominantly $W^3$. 
As $\delta_B$ is increased to $20$ the couplings of the two states become
comparable. 
However, the light state still makes a negligible contribution to the part
of the amplutude responsible for PUV. To see this, note that the PUV can be
traced to incomplete cancellations in the term that grows linearly with s at
high energies. The contribution of the $k$th state to this sum rule is
proportional to $m_k^2 g_{11k}^2$, so the light state has little effect.
In the case where $\delta_D$ is non-vanishing, it is the other member
of the degenerate KK pair whose mass is lowered.  In this case, the
light state then couples mostly to isospin, and is capable of
significantly modifying the $W_L^+W_L^-$ scattering amplitude.

%
\nn

A note about numerical instabilities is in order. We find that the sum rule
governing the coefficient of the $s^2$ term is satisfied at the level
of $10^{-6}$ after the first KK state is included, while the sum 
rule for the $s$ term is satisfied to
the level of $10^{-2}$. After a few more states are included, 
the first sum rule is satisfied to ${\cal O}(10^{-9})$,
and the second to ${\cal O}(10^{-3})$. 
This demonstrates that the PUV is due to incomplete
cancellations in the term growing like $s$,
as well as the presence of the constant term. However, consider the
case where there is a
numerical instability in the calculation of the 
couplings at the $10^{-8}$ level. Then we can
estimate the energy scale at which this becomes important by noting that the
amplitude goes like $ 10^{-8}(s^2/M_W^4)$. This becomes of order unity when
$\sqrt{s} \sim 8 \tev$, implying that a calculation good to only 8 digits will
give incorrect results when the scale of PUV is in the few $\tev$ range. 
Since unitarity depends on delicate cancellations, it could be expected
that any error will decrease the scale of PUV. However, we have, 
somewhat surprisingly, found that this is {\it not} necessarily true. The
reason is that the terms growing like $s$ and $s^2$ have the opposite sign. A
numerical error can thus cause the $s^2$ term to turn on prematurely and
cancel the contribution from the $s$ term, leading to an
apparent scale of PUV {\it higher} than it actually is. 
For example, we studied one case where a
numerical error at the level of $10^{-8}$ 
caused the apparent scale of PUV to be
$12 \tev$, while the correct scale was actually $6 \tev$.   For this
reason, all our results were
computed independently on two platforms, with agreement to greater than 12
digits.

\section{Conclusions}

The Warped Higgsless Model, which breaks the electroweak symmetry
via boundary conditions associated with an extra dimension,
offers a promising alternative to the Higgs mechanism.
A custodial $SU(2)_D$ symmetry is present in the model, so that
reasonable agreement with  precision electroweak
data is conceivable.  However, the degree of such agreement varies as the
parameter space of the model is explored, and some regions can
be excluded.    Here, we examined the effects of including
IR(TeV)--brane kinetic brane terms associated with the $U(1)_{B-L}$
and $SU(2)_D$ gauge symmetries of the model.  We found that the 
addition of the $U(1)_{B-L}$ kinetic term enhances the agreement with the
tree-level SM electroweak relations, particularly for larger values of the
ratio of the 5--$d$ couplings $g_{5R}/g_{5L}$, with reasonable values
of the brane term parameter $\delta_B$.  However, including the 
$SU(2)_D$ brane term alone
results in a stark disagreement with the SM tree level relations in
the electroweak sector.
We performed a limited exploration of the full parameter space and
found it is possible that a combination of the two IR--brane terms may
result in a reasonable consistency with the tree-level SM relations.

In its original form, the WHM has some difficulties
in the gauge sector in that perturbative unitarity is violated
at the TeV-scale in $W_L^+W_L^-$ scattering.  This does not exclude
the model from being viable, but does suggest that interactions
in the gauge sector are problematic.
To restore unitarity in the gauge sector, additional new physics
must be introduced.  Here, we again examined the effects of including the
IR(TeV)--brane kinetic brane terms.  It is well-known that the addition
of brane terms can alter the couplings and masses of gauge KK
states, and this would thus affect the KK contributions to $W_L^+W_L^-$
scattering.  While we found that the $U(1)_{B-L}$ brane term
does modify the gauge KK spectrum, we also discovered that PUV in
$W_L^+W_L^-$ scattering is independent of such a brane term and
hence remains unaffected by its presence.  This is because this
scattering process is inherently non-Abelian and should
not depend on the $U(1)_{B-L}$ dynamics.  In contrast, the inclusion of
the $SU(2)_D$ kinetic term does affect $W_L^+W_L^-$ scattering; for
moderate values of the brane term, violation of perturbative unitarity 
is delayed until $\sqrt s\approx 7$ TeV.
In addition, we also investigated the collider bounds on the production
of the lightest gauge KK excitation as a function of the brane terms.
Searches for new gauge bosons at the Tevatron and LEP II exclude
large values of the kinetic term parameters $\delta_{B,D}$.

Our analysis shows how various directions in the parameter space of
the WHM affect its phenomenology.  Requiring a model that is
perturbatively sensible up to ${\cal O}(10)$ TeV favors $\kappa=1$
and $1 \lsim \delta_D \lsim 10$, regardless of the size of
$\delta_B$.  Collider constraints on the KK modes of the gauge bosons
can accommodate this range of parameters, as long as $\delta_{B,D}
\lsim 2-3$, with the Tevatron bounds depending on the fermion localization.  
We observe that the requirements of multi-TeV
perturbative unitarity and those imposed by tree level SM relations,
as represented by the pseudo-oblique parameters and various values of
$\sin^2 \theta$, do not coexist without tension in this model.
However, a direct comparison of these latter quantities with the
electroweak data requires a computation of the radiative corrections
in the WHM, which lies outside the scope of this work.  Thus, it
remains an open possibility that this model could provide a viable
alternative for electroweak symmetry breaking, valid far above the
weak scale.

\noindent{\Large\bf Acknowledgements}

We would like to thank Csabi Csaki, Graham Kribs, Hitoshi Murayama, and
John Terning for discussions related to this work.
The work of H.D. was supported by the Department of Energy under
grant DE-FG02-90ER40542.

%
\def\MPL #1 #2 #3 {Mod. Phys. Lett. {\bf#1},\ #2 (#3)}
\def\NPB #1 #2 #3 {Nucl. Phys. {\bf#1},\ #2 (#3)}
\def\PLB #1 #2 #3 {Phys. Lett. {\bf#1},\ #2 (#3)}
\def\PR #1 #2 #3 {Phys. Rep. {\bf#1},\ #2 (#3)}
\def\PRD #1 #2 #3 {Phys. Rev. {\bf#1},\ #2 (#3)}
\def\PRL #1 #2 #3 {Phys. Rev. Lett. {\bf#1},\ #2 (#3)}
\def\RMP #1 #2 #3 {Rev. Mod. Phys. {\bf#1},\ #2 (#3)}
\def\NIM #1 #2 #3 {Nuc. Inst. Meth. {\bf#1},\ #2 (#3)}
\def\ZPC #1 #2 #3 {Z. Phys. {\bf#1},\ #2 (#3)}
\def\EJPC #1 #2 #3 {E. Phys. J. {\bf#1},\ #2 (#3)}
\def\IJMP #1 #2 #3 {Int. J. Mod. Phys. {\bf#1},\ #2 (#3)}
\def\JHEP #1 #2 #3 {J. High En. Phys. {\bf#1},\ #2 (#3)}

\end{document}